\documentclass[12pt]{article}
\usepackage[centertags]{amsmath}
\usepackage{amssymb}
\usepackage{amsthm}
\usepackage{graphicx}

\usepackage{cite}

\newcommand{\be}{\begin{equation}}
\newcommand{\ee}{\end{equation}}
\newcommand{\ba}{\hspace*{-5pt}\begin{array}}
\newcommand{\ea}{\end{array}}
\newcommand{\p}{\partial}

\usepackage[pagewise,mathlines]{lineno}

\numberwithin{equation}{section}

\newtheorem{proposition}{Proposition}

\newtheorem{theorem}{Theorem}

\theoremstyle{definition}
\newtheorem{remark}{Remark}
\newtheorem{definition}{Definition}

\newcommand{\fr}[2]{\tfrac{#1}{#2}}
\newcommand{\pd}[2]{\frac{\partial#1}{\partial#2}}

\renewcommand{\thefootnote}{\fnsymbol{footnote}}

\renewcommand{\ba}{\begin{array}}
\renewcommand{\ea}{\end{array}}
\newcommand{\beg}{\begin{eqnarray}}
\newcommand{\eeq}{\end{eqnarray}}
\newcommand{\bg}{\begin{eqnarray*}}

\newcommand{\ed}{\end{eqnarray*}}

\newcommand{\nn}{\nonumber}

\renewcommand{\p}{\partial} 
 
\newcommand{\notlhd}{\lhd\kern-.8em{/}\ } 
\newcommand{\notexist}{\ \exists\kern-.5em{\raise.1em\hbox{/}}\ }

\newcommand{\pde}[2]{\frac{\p #1}{\p #2}}

\newcommand{\inp}{{\mbox{\vbox{\hrule width0ex\hbox{\vrule
 height0ex\kern3.8pt
\vbox{\kern2.5pt}\kern3.8pt \vrule height1.6ex}
\hrule width1.6ex}}}}

\title{Compacton equations and integrability: the Rosenau-Hyman and Cooper-Shepard-Sodano
equations} 
\author{R Hern\'andez Heredero$^1$, M Euler$^2$, N Euler$^2$ and E G Reyes$^3$\\ \smallskip \\
\small{$^1$ Departamento de Matem\'atica Aplicada a las TIC}\\
\small{Universidad Polit\'ecnica de Madrid. C.~Nikola Tesla s/n. 28031 Madrid. Spain}\\
\small{$^2$ Division of Mathematics,
Department of Engineering Sciences and Mathematics}\\  
\small{Lule\aa\ University of Technology, SE-971 87 Lule\aa, Sweden}\\
\small{$^3$ Departamento de Matem\'atica y Ciencia de la Computaci\'on}\\
\small{Universidad de Santiago de Chile, Casilla 307 Correo 2, Santiago, Chile}\\
}
\date{April 2, 2019}
\begin{document}

\maketitle

\begin{abstract}
We study integrability --in the sense of admitting recursion operators-- of two nonlinear equations which are known to 
possess compacton solutions: the $K(m,n)$ equation introduced by Rosenau and Hyman 
  \[ D_t(u) + D_x(u^m) + D_x^3(u^n) = 0 \; , \]
and the $CSS$ equation introduced by Coooper, Shepard, and Sodano,
  \[ D_t(u) + u^{l-2}D_x(u) + \alpha p D_x (u^{p-1} u_x^2) + 2\alpha D_x^2(u^p u_x) = 0 \; . \]
We obtain a full classification of {\em integrable $K(m,n)$ and $CSS$ equations}; we present their recursion operators,  and we prove that all of them
are related (via nonlocal transformations) to the Korteweg-de Vries equation. As an application,
we construct isochronous hierarchies of equations associated to the integrable cases of $CSS$.
\let\thefootnote\relax
\let\thefootnote\relax
\end{abstract}

\section{Introduction}
We begin by quoting Rosenau \cite{Ro}: `'We define a {\em compact wave} as a robust solitary wave with {\em compact support} beyond which it vanishes identically. We then define a {\em compacton} as a compact wave that preserves its shape after interacting with other compacton''. Rosenau and Hyman found examples of compactons while studying generalizations of the Korteweg-de Vries equation for which the dispersion term 
is nonlinear. Their model equation is the so-called $K(m,n)$ equation
\begin{equation}\label{rhintro}
u_t + (u^m)_x + (u^n)_{xxx} = 0 \: ,
\end{equation}
and an example of a compacton bearing equation within the family \eqref{rhintro} is $K(2,2)$. 
In this case, the function $u(x,t) = (4 c/3) \cos^2((x-c t)/4)$ for $|x-c t| \leq 2 \pi$ and $u(x,t) = 0$ otherwise, is a compacton solution. Further works on compactons are~\cite{LD,LOR} and the comprehensive review \cite{RZ}. 

It turns out that solutions to equations within the $K(m,n)$ can exhibit very complex behaviors; we refer the reader to \cite{AW,AW2,ASWY}, and to the papers \cite{LOR,RZ,ZR} authored by Rosenau and his coworkers, for general discussions. Here,
we just mention one example:  in \cite{RH} the authors present four local conservation laws of $K(2,2)$, and credit P.J. Olver with the observation that
no further local conservation laws seem to exist\footnote{This observation has been proven rigorously by Vodov\'a
in 2013, see \cite{Vo}.}. This (non)existence of conservation laws has an important analytic implication, see \cite{ZR}: 
initially nonnegative, smooth and compactly supported solutions to~$K(m,n)$ lose their smoothness within a finite time.  

We wonder if this complex behavior has to do with (lack of) integrability. In this work we present a detailed study of the integrability properties of $K(m,n)$. We find that, module a rather general space of allowable transformations, the only integrable
equations belonging to the $K(m,n)$ family are the KdV and modified KdV equations, and that integrable equations within the $K(m,n)$ family cannot have compacton solutions. In particular, we recover the observation in \cite{HHR,Vo} that~$K(2,2)$ is not integrable. 

In order to obtain this result we classify all integrable $K(m,n)$ equations using the theory of formal symmetries (to be summarized in Section 2). The power of this approach has been amply demonstrated by the classification results for evolution equations and systems of equations due to researchers
such as Shabat, Fokas, Svinolupov, Sokolov, Mikhailov and others (see \cite{MSS,Fok,SviSok,HSSv,Her1,Her2}), and also
by the important papers \cite{SW,SW1} on the 
classification of integrable scalar evolution equations satisfying an homogeneity condition.

Since our search for integrable compacton bearing equations within the~$K(m,n)$ class does not yield  examples, we also investigate a related family, the Cooper-Shepard-Sodano family of equations
\begin{equation}
u_{t}+u^{l-2}u_{x}-\alpha pD_{x}\left(u^{p-1}u_{x}^{2}\right)+2\alpha D_{x}^{2}\left(u^{p}u_{x}\right)=0,                                                        \qquad\alpha\neq0 \; ,  \label{cssintro}
\end{equation}
introduced in \cite{CSS}. We quote from this paper: ``These equations have the same terms as the equations considered by Rosenau and Hyman, but the relative weights of the terms are quite different leading to the possibility that the integrability properties might be different''. The authors of \cite{CSS} then proceed to show that their family of equations indeed admits compacton bearing equations. One such  equation is~\eqref{cssintro} with $l=3$, $p=2$. This family of equations is further studied in \cite{KC,DK}.

Encouraging properties of (\ref{cssintro}) are the facts that it admits a Hamiltonian formulation, and that it
possesses three physically interesting conservation laws: area, mass and energy. Regretfully, we prove herein that they are not integrable in general. Using formal symmetries once more, we obtain {\em six} integrable equations within the~(\ref{cssintro}) family. None of them can support compacton solutions.

Since the existence of compacton solutions {\em is} a rather extraordinary occurrence in the nonlinear world,
we believe that our results are not only  important by themselves, but also because they seem to express certain
rigidity in our present algebraic/geometric/analytic approach to integrability. In other words, $K(2,2)$ say, must
be ``special'', and so far we have not been able to uncover the deeper source of its special character.

Our paper is organized as follows. We review the theory of formal symmetries and integrability in Section 2
after, essentially, \cite{MSS}, and in Section 3 we use this theory to classify integrable $K(m,n)$ equations.
We note that a previous classification has appeared in \cite{HHR}. One integrable case was missing therein
and we single it out here. Fortunately, the missing case does not alter the conclusion in \cite{HHR} that the only~$K(m,n)$ integrable case are (essentially, module a class of allowable transformations specified in Section 3) the KdV and mKdV equations. The present classification also differs from the one appearing in  \cite{HHR} in that here
we explain in detail how to connect our integrable cases to KdV (or, to the linear equation) and because in Section 5 we
exhibit explicit recursion operators for all our integrable $K(m,n)$ equations. In Section 4 we study integrability of the Cooper-Shepard-Sodano family and again we are able to explain how to connect its integrable cases to KdV
(or, to the linear equation), and to exhibit recursion operators. Finally in Section 6 we present an application of our results: we construct integrable isochronous equations, after 
\cite{Calogero-2005,Calogero-Mariani-2005,Calogero-Euler-Euler}, starting from the equations in our classification of integrable CSS equations, explain how to obtain their point symmetries, and present their corresponding recursion operators.



\section{Formal symmetries and integrability}

The formal symmetry approach to integrability~\cite{MSS,MikSok} begins with the observation that standard (systems 
of) partial differential equations which are integrable (for instance, in the sense of Calogero, see \cite{Ca}) 
usually admit an infinite set of (generalized) symmetries of arbitrarily large differential order. A.B. Shabat and 
his collaborators, see for instance \cite{MSS,MikSok}, realized that it is possible to weaken the notion of a 
(generalized) symmetry to the notion of a {\em formal} symmetry ---to be defined precisely below---
and that this new concept provides a computationally efficient tool for defining integrability and classifying 
integrable equations. We recall 
from \cite{MSS,O} that~$G = (G^\alpha)$ is a symmetry of a system of partial differential equations of the form 
$\Delta_a(x^i,u^\alpha,u^\alpha_{x^i},\dots) =0$,
if
\begin{equation}\label{genSym}
\Delta_*(G)=0 \; 
\end{equation}
whenever $u^\alpha(x^i)$ is a solution to $\Delta_a=0$, where $\Delta_*$ is the formal linearization of the system 
$\Delta_a=0$, that is,
$\Delta_*=\left( \sum_{L}\frac{\partial\Delta_a}{\partial u_L^\alpha}D_L \right)$. If the system $\Delta_a=0$
consists of just one scalar evolution equation,
\begin{equation} \label{de}
 \Delta=u_t-F \; ,
\end{equation}
then equation~\eqref{genSym} becomes $D_t G=F_*(G)$ or, equivalently,
\begin{equation}\label{evolSym}
D_tG = D_\tau F \; ,
\end{equation}
where
\[
{D}_{\tau} = \sum_{\# K \geq 0} {D}_{K}(G) \,\frac{\partial\ }{\partial u_{K}}  \; .
\]

\smallskip

{\em Note: Here and henceforth we use standard notation from the geometric theory of differential equations as presented in {\rm \cite{O}}, see also {\rm \cite{MSS}}.}

\medskip

Following \cite{MSS,MikSok}, we apply a second linearization to formula \eqref{evolSym}.
We obtain, using some formulae appearing in \cite{MSS},
\begin{equation} \label{dtaulocal2}
(D_tG)_*=(D_\tau F)_*\quad\Leftrightarrow\quad D_t(G_*)+G_*\circ F_*=D_\tau(F_*)+F_*\circ G_* \; ,
\end{equation}
in which $D_t ( \sum_{L} a_L D_L ) =  \sum_{L} D_t( a_L) D_L$ if 
$G_\ast = \sum_{L} a_L D_L$, and the last equality holding on solutions to \eqref{de}. The expression~$D_\tau(F_*)$ is defined 
analogously. 
We interpret our symmetry condition (\ref{dtaulocal2}) using commutators:
\begin{equation}\label{symOp}
D_t(G_*)-[F_*,G_*]=D_\tau(F_*) \; .
\end{equation}

Let us consider the degree of the operators appearing in \eqref{symOp}. The
degree of~$F_*$ as a differential operator ---let us denote it by~$\deg(F_*)$--- is the differential
order of~$F$, i.e.~the order of the differential equation \eqref{de} and thus, it is fixed. The degree of the
left hand side of~\eqref{symOp} depends on $G$: $D_t(G_*)-[F_*,G_*]$ is a differential operator generically of degree 
$\deg(G_*$) plus~$\deg(F_*)$ minus 1, much higher than that of the operator in the right hand side, 
of degree $\deg(F_*)$, if there are high order symmetries~$G$. Thus, it is not clear at all that non-trivial solutions to \eqref{symOp} should exist: the existence of (generalized) 
symmetries~$G$ of arbitrarily high differential order must impose extremely strong constraints on the function $F$. 

Following \cite{MSS,MikSok}, and partially motivated by the theory of recursion operators, see \cite{O}, we define
formal symmetries using the left hand side of Equation \eqref{symOp}:

\begin{definition} \label{fs}
Let $u_t=F$ be an evolution equation with~$F$ a function of two independent variables~$x$, $t$,
one dependent variable~$u$ and a finite number of derivatives of~$u$ with respect to~$x$.
A formal symmetry of rank~$k$ of this partial differential equation is a formal pseudo-differential
operator
\begin{equation} \label{pdo}
\Lambda=l_rD^r+l_{r-1}D^{r-1}+\cdots+l_0+l_{-1}D^{-1}+l_{-2}D^{-2}+\cdots,\qquad D=D_x
\end{equation}
with~$l_i$ being functions of $t$, $x$, $u$ and finite numbers of $x$-derivatives of~$u$,
that satisfies the equation
\begin{equation}\label{formSym}
D_t(\Lambda)=[F_*,\Lambda]
\end{equation}
whenever $u$ is a solution to $u_t =F$, up to a pseudo-differential operator of degree~$r+\deg(F_*)-k$.
A formal symmetry of infinite rank is a pseudo-differential operator \eqref{pdo} such that
\eqref{formSym} holds identically whenever $u$ is a solution to $u_t =F$.
\end{definition}

Note that if~$G$ is a symmetry of order~$p$ of~$u_t=F$, then~\eqref{symOp} implies that~$G_*$
is a formal symmetry of rank~$p$. We also remark that a formal symmetry of infinite rank
is a recursion operator, see \cite{O}. Thus, it generates, in principle,
an infinite number of generalized symmetries of the equation at hand. For example, see \cite{DS}, it can be proven that 
application of a {\em quasilocal} recursion operator (in the sense of \cite[Section 1]{DS}) to a given symmetry yields a (generalized) symmetry, and so  
such an operator could indeed generate an infinite chain of (generalized) symmetries.

The main technical point behind Definition \ref{fs} is that the space of solutions of equation~\eqref{formSym} 
is much richer and structured  than that of equation~\eqref{symOp} or even~\eqref{genSym}. For example, powers and 
roots of formal symmetries (computed using the standard theory of formal pseudo-differential operators, see 
\cite{MSS,O}) are also formal symmetries. In fact, this observation was one of the original motivations for the 
use of formal pseudo-differential operators in Definition \ref{fs},
because the $r$th~root of a differential operator~\eqref{pdo} is usually a {\em pseudo}-differential operator. 

Now we explain why Definition \ref{fs} restricts the function $F$. A theorem due to M. Adler, see \cite{MAdler},  states 
that the residue (the coefficient of~$D^{-1}$) of a commutator of formal pseudo-differential operators is always a 
total derivative. If we apply this result to different powers~$\Lambda^{i/r}$ of a generic formal symmetry
\footnote{The foregoing discussion implies that instead of a general formal 
symmetry~$\Lambda$ of degree~$r$, we can consider its $r$th root~$\Lambda^{1/r}$ of degree~1 without loss of 
generality, see \cite{MSS} for details.} $\Lambda$ of rank~$k$ inserted into~\eqref{formSym}, we obtain 
\[
D_t (\operatorname{residue} (\Lambda^{i/r})) = 
\operatorname{residue}{D_t(\Lambda^{i/r})}= \operatorname{residue}[F_*,\Lambda^{i/r}]=D\sigma_i 
\]
for some differential functions $\sigma_i$, i.e.~a sequence of conservation laws 
\begin{equation}\label{eq:ccl}
D_t\rho_i\doteq D_x\sigma_i,\qquad i=-1,1,\ldots,
\end{equation}
which are, together with the special case~$D_t\rho_0=D_t(l_r/l_{r-1})=D_x\sigma_0$, the so called~\emph{canonical 
conservation laws}. 
The symbol~$\doteq$ means that equations~\eqref{eq:ccl} must hold on solutions of~\eqref{de}, i.e.~all derivatives with respect to~$t$ must be substituted using the equation and its 
differential consequences. 

As observed in \cite{MSS}, the {\em canonical densities}~$\rho_i$ and conserved fluxes~$\sigma_i$ are differential
functions which can be recursively written in terms 
of the right hand side~$F$ of the equation and its derivatives. The fact that the left hand side 
of~\eqref{eq:ccl} must be a total derivative with respect to~$x$ for all $i =-1,1,2\cdots$, produces obstructions 
that are 
necessary conditions for the existence of (generalized/formal) symmetries~$G$, i.e.~for~integrability.

For example, for evolution equations of third order, 
\begin{equation}\label{eq:eq3}
u_t=F(x,u,u_x,u_{xx},u_{xxx}) \; ,
\end{equation}
the first canonical density is 
$\rho_{-1}=\left(\partial F \big/\partial u_{xxx}\right)^{-1/3}$, see \cite{MSS}. Therefore, a first integrability 
condition is requiring~$D_t\rho_{-1}$ to be the total derivative of a local function~$\sigma_{-1}$. The second 
canonical density imposes further differential restrictions on~$F$, and so forth.
Usually, after a small number of steps our family of equations either fails to satisfy the integrability conditions, 
or the right hand side~$F$ becomes so specific that we are able to produce a formal symmetry of infinite rank and 
therefore, in principle, a sequence of generalized symmetries of $u_t = F$. If $u_t =F$ represents a family of 
equations, this procedure allows us to find all integrable cases in the family. 
We are led to the following precise definition of integrability, after \cite{MSS,O}:

\begin{definition} \label{int}
A system of evolution equations is integrable if and only if it
possesses a formal symmetry of infinite rank.
\end{definition}

Let us we write down the first five canonical densities  for a third order equation~\eqref{eq:eq3} 
following~\cite{MSS}. We will use them in the next section to study the integrability of the Rosenau-Hyman and 
Cooper-Shepard-Sodano equations:

\begin{proposition} \label{p1}
Let $u_t=F(x,u,u_x,u_{xx},u_{xxx})$ be an arbitrary third order evolution equation. The first five canonical 
conserved densities can be written explicitly as
\begin{linenomath}
\begin{align}
\rho_{-1}&=\left(\pd{F}{u_{xxx}}\right)^{-1/3},\label{ic0}\\
\rho_0&=\rho_{-1}^3\pd{F}{u_{xx}},\\
\rho_1&=D_x\left(2\rho_{-1}^{-2}u_{x}+\rho_{-1}^2\pd{F}{u_{xx}}\right)+
     \rho_{-1}^{-3}(D_x\rho_{-1})^2+{\frac{1}{3}}\rho_{-1}^5\left(\pd{F}{u_{xx}}\right)^2\nonumber\\
&\qquad  {}+\rho_{-1}(D_x\rho_{-1})\pd{F}{u_{xx}}
    -\rho_{-1}^2\pd{F}{u_{x}}+\rho_{-1}\sigma_{-1},\\
\rho_2&={}-{\frac{1}{3}}(D_x^2\rho_{-1})\pd{F}{u_{xx}}-(D_x\rho_{-1})\pd{F}{u_{x}}+
     \rho_{-1}\pd{F}{u}+\rho_{-1}^{-1}(D_x\rho_{-1})^2\pd{F}{u_{xx}}\nonumber \\
&\qquad  {}-{\frac{1}{3}}\rho_{-1}^4\pd{F}{u_{x}}\pd{F}{u_{xx}}
    +{\frac{1}{3}}\rho_{-1}^3(D_x\rho_{-1})\left(\pd{F}{u_{xx}}\right)^2\nonumber \\
&\qquad  {}+{\frac{2}{27}}\rho_{-1}^7\left(\pd{F}{u_{xx}}\right)^3+{\frac{1}{3}}\rho_{-1}\sigma_{0}, 
    \\
\rho_3&=\rho_{-1}\sigma_{1}-\rho_{1}\sigma_{-1}.\label{ic4}
\end{align}
\end{linenomath}
\end{proposition}

\begin{remark}
The condition of existence of a formal symmetry of infinite rank is strictly {\em weaker} than
the condition of existence of an infinite number of generalized symmetries. Indeed, the $2$ component
system
\[\left \{
{\setlength\arraycolsep{2pt}
\begin{array}{rl}
u_t =& u_{xxxx} + v^2 \\
v_t =& v_{xxxx} \; ,
\end{array}}
\right.
\]
considered by Bakirov in \cite{Bakirov},
possesses exactly {\em one} generalized symmetry, as proved by Beukers, Sanders and Wang
\cite{BSW}. On the other hand, it does possess a recursion operator \cite{Bilge}.
\end{remark}

\smallskip

Now, a very important remark, see \cite{MSS} and also \cite{O}, is that the use of transformations between equations
is a very convenient way to proceed when seeking classifications. 
In this paper we deal with integrable equations of the type
\begin{equation}\label{eq:equfu}
u_t=f(u)u_{xxx}+g(u,u_x,u_{xx}) \; ,
\end{equation}
see \cite{RH,CSS}. The general strategy we use to classify these equations consists in performing a sequence of 
convenient point transformations and differential substitutions that preserve integrability, and then apply 
and compute the integrability conditions associated to~\eqref{ic0}--\eqref{ic4}. Our general procedure is as follows:

First, if $f'(u)\neq0$ the point 
transformation~$u\to f(u)^{-1/3}$ converts Equation~\eqref{eq:equfu} into another one of the form
\begin{equation}\label{equdu3}
u_t=D_x\left[\frac{u_{xx}}{u^3}+f_1(u,u_x)\right]+f_2(u,u_x,u_{xx}) \; , 
\end{equation}
where~$f_2(u,u_x,u_{xx})$ {\em is not} a total $x$-derivative. This form is very convenient because it follows from 
(\ref{ic0}) that the integrability condition~$D_t\rho_{-1}=D_x\sigma_{-1}$ is equivalent to requiring that~$f_2=0$.
This condition greatly restricts the form of the equation. Once~$f_2=0$, the equation admits a potentiation~$u\to u_x$ that brings it into the form~$u_t=u_{xxx}/u_x^3+f_1(u_x,u_{xx})$. A subsequent hodograph transformation~$x\to u$, $u\to x$ simplifies it to one of the form
\[
u_t=u_{xxx}+h(u_x,u_{xx}).
\]
This equation can be ``antipotentiated" ($u_x\to u$) to get
\[
u_t=u_{xxx}+D_xh(u,u_x).
\]
Our integrability conditions imply that the integrable cases of this equation are all of the form
\begin{equation}\label{equdx}
u_t=u_{xxx}+D_x\left[h_2(u) u_{x}^2+\left(a+bu\right)u_x+h_0(u)\right].
\end{equation}
If~$h_2(u)\neq0$ a further point transformation~$\int\!\exp\left[\frac23\int\!{h_2(u)\,du}\right]\,du\to u$ 
transforms this equation into another one of the form
\begin{equation}\label{equcdff}
u_t=u_{xxx}+(a+bu)\,u_{xx}+f_1(u)\,u_x^3+f_2(u)\,u_x^2+f_3(u)\,u_x.
\end{equation}

We will show that all the integrable cases of the RH and CSS equations can be written in the form~\eqref{equcdff}, 
as linear equations, KdV or mKdV, or the Calogero-Degasperis-Fokas (CDF) equation (see below; the CDF is 
Miura-transformable to KdV). Thus, if 
we ``pullback" the recursion operator of KdV (or, the linear equation) by the foregoing transformations, we can 
construct recursion operators of the original equations, and therefore we obtain an explicit proof of integrability
in terms of Definition \ref{int}.  In actual fact, we seldom perform this pullback operation explicitly.
Once we know that a given equation is integrable, it is usually straightforward to compute its recursion operator
from first principles, as in \cite{PNN}.

Now we carry out this plan.

\section{Integrability of the Rosenau-Hyman equation}

We consider the compacton equation of Rosenau and Hyman (see \cite{RH})
\begin{equation}\label{eq:eqRH}
D_t(u)+D_x(u^m)+D_x^3(u^n)=0,\qquad n\neq0 \; .
\end{equation}
As we informed in Section 2, $D_t$ and $D_x$ are total derivatives with respect to independent variables~$t$ and~$x$. For simplicity, we 
use the subindex notation~$u_t=D_tu$, $u_x=D_xu$, $u_{xx}=D_x^2$ but we prefer to use the total derivative notation 
when applied to a more complicated differential function, e.g.~$D_x(u^m)=mu^{m-1}u_x$.

The case~$n=1$ i.e.
\[
u_t+mu^{m-1}u_x+u_{xxx}=0 \; ,
\] 
is well-known, see \cite[Section 4.1]{MSS}: the  only integrable cases are~$m=0,1,2,3$, i.e.~the linear equation, 
the KdV and the modified KdV equations. We write them as (using the point transformation $x\to -x$)
\begin{linenomath}
\begin{align}
u_t&=u_{xxx}+\alpha\, u_x+\beta,\quad\alpha,\beta\in\mathbf{C}, \label{34} \\
u_t&=u_{xxx}+2\,u u_x, \label{35} \\
u_t&=u_{xxx}+3\,u^2u_x. \label{36}
\end{align}
\end{linenomath}

If $n\neq1$, the point transformation~$x\to-x$, $t\to t/n$, $u\to u^{3/(1-n)}$ changes~\eqref{eq:eqRH} into 
equations of the form~\eqref{equdu3}, namely:
\begin{equation}\label{eq:eq1}
u_{t}=D_{x}\left[\frac{u_{xx}}{u^3}
-\frac{m (n-1) }{n (3 m-n-2)}u^{\frac{2-3 m+n}{-1+n}}
-\frac{9 n}{2 (n-1)}\frac{u_{x}^2}{u^4}
\right]
+\frac{(n+2) (2 n+1) u_{x}^3}{(n-1)^2 u^5}
\end{equation}
if $3m-n-2\neq0$, and 
\begin{equation}\label{eq:eq2}
u_{t}=D_{x}\left[\frac{u_{xx}}{u^3}-\frac{9 n}{2 (n-1)} \frac{u_{x}^2}{u^4}
+\frac{(n+2)}{3 n}\log{u}\right]
+\frac{(n+2) (2 n+1) u_{x}^3}{(n-1)^2 u^5}
\end{equation}
if $3m-n-2=0$.

The first integrability condition $D_t\rho_{-1} = D_x \sigma_{-1}$ or, equivalently, 
$u_t\in\operatorname{Im}D_x$ implies that either~$n=-2$ or~$n=-1/2$ in both cases, because the last term 
in~\eqref{eq:eq1} and~\eqref{eq:eq2} must be zero. The composition of a potentiation, a hodograph and an 
antipotentiation yields the following equations of the form~\eqref{equdx}:
\begin{equation}\label{eq:eq1c}
u_t=u_{xxx}+D_x\left[\frac32\frac{n+2}{n-1}\frac{u_{x}^2}{u} + 
    \frac{m (n-1)}{n (3 m-n-2)}\,u^{\frac{3 (m-1)}{n-1}}\right],
\end{equation}

\begin{equation}\label{eq:eq2c}
u_t=u_{xxx}+D_x\left[\frac32 \left( \frac{n+2}{n-1} \right) \frac{u_{x}^2}{u}+\frac{n+2}{3 n}\,u \log{u}\right].
\end{equation}

If~$n=-2$, Equation \eqref{eq:eq1c} becomes
\[
u_t=u_{xxx}-\frac{m-1}{2} u^{-m}u_{x}
\]
whose integrable cases are again~$m=1,0,-1,-2$, corresponding to linear equations, KdV and mKdV.  
On the other hand, if~$n=-2$, Equation \eqref{eq:eq2c} becomes a linear equation included in case~\eqref{34}.

If~$n=-1/2$, Equation \eqref{eq:eq1c} can be written in the form~\eqref{equcdff}, this is,
\[
u_t=u_{xxx}-\frac12u_{x}^3-\frac{4 m (m-1)}{2 m-1}\, {\rm e}^{(1-2 m)u} u_{x} \; ,
\] 
after a point transformation~$u\to{\rm e}^{u}$. This family of equations satisfies the first two integrability 
conditions. The third integrability condition is~$D_t\rho_1\in\operatorname{Im}D_x$, and the canonical conserved
density~$\rho_1$ satisfies
\[
D_t(\rho_1)\sim -2 (m-1) m (2 m-3) (2 m+1) e^{u-2 m u} u_{x}^3 \; , 
\]
in which the symbol~$\sim$ denotes equality except for the addition of a total $x$-derivative. Thus, integrability
can be achieved only in the cases~$m=-1/2$, $0$, $1$, $3/2$\footnote{We note that the case $n=-1/2$, $m=3/2$ was 
missing in~\cite{HHR}.}
which are all subcases of the Calogero-Degasperis-Fokas (CDF) equation
\begin{equation} \label{cdf}
u_t=u_{xxx}-\frac12u_x^3+\left(\alpha{\rm e}^{2u}+\beta{\rm e}^{-2u}+\gamma\right)u_x\; .
\end{equation}
Finally, when $n=-1/2$ Equation~\eqref{eq:eq2c} becomes
\[
u_t=u_{xxx}-3\frac{u_{x} u_{xx}}{u}+\frac32\frac{u_{x}^3}{u^2}-(1+\log{u})u_{x}
\]
and for it
\[D_t(\rho_1)\sim-2\frac{u_x^3}{u^3}
\]
so this case is not integrable.

\begin{remark}
We note that the CDF equation~\eqref{cdf} can be related to the KdV equation through the Miura transformation 
\[
\frac32 u_{xx}-\frac34 u_{x}^2 -\sqrt{6 \beta } {\rm e}^{-u} u_{x}-\frac{1}{2} \alpha  e^{2 u}-
\frac{1}{2} \beta  e^{-2 u}-\frac{\gamma }{2} \to u\; .
\]
\end{remark}

\smallskip

We summarize the integrable cases of the Rosenau-Hyman family~\eqref{eq:eqRH} in the following theorem. 
We make the point transformation~$x\to-x$, $t\to t/n$ and write
\begin{equation}\label{eqRH1}
u_t=\tfrac{1}{n}D_x(u^m)+\tfrac{1}{n}D_x^3(u^n)\; ,\qquad n\neq0 
\end{equation}
instead of \eqref{eq:eqRH}.
This transformation is invertible and does not affect integrability.
\newpage
\begin{theorem}
The integrable cases of the Rosenau-Hyman family \eqref{eqRH1} 
are
\begin{enumerate}
\item $n=1$, $m=0,1,2,3$, corresponding to Equations \eqref{34}, \eqref{35} and \eqref{36}, namely,
\begin{linenomath}
\begin{align}
u_t&=u_{xxx}+\alpha\, u_x+\beta,\quad\alpha,\beta\in\mathbf{C}, \label{341} \\
u_t&=u_{xxx}+2\,u u_x, \label{351} \\
u_t&=u_{xxx}+3\,u^2u_x \; . \label{361}
\end{align}
\end{linenomath}
\item $n=-2$, $m=-2,-1,0,1$, corresponding to Equations 
\begin{linenomath}
\begin{align}
u_t&=D_x\left[D_x\left(\frac{u_x}{u^{3}}\right)-\frac{1}{2u^2}\right] \; , \label{eq:eqm2m2} \\
u_t&=D_x\left[D_x\left(\frac{u_x}{u^{3}}\right)-\frac{1}{2u}\right] \; , \label{39} \\
u_t&=-\frac12D_x^3\left[ u^{-2} \right]
    \; ,\label{395} \\
u_t&=D_x\left[D_x\left(\frac{u_x}{u^{3}}\right)-\frac{1}{2}u\right]\label{310}
\end{align}
\end{linenomath}
respectively.
\item $n=-\fr12$, $m=\fr32,1,0,-\fr12$, corresponding to Equations 
\begin{linenomath}
\begin{align}
u_t&=D_x\left[D_x\left(\frac{u_x}{u^{3/2}}\right)-2u^{3/2}\right], \label{312} \\
u_t&=D_x\left[D_x\left(\frac{u_x}{u^{3/2}}\right)-2u\right], \label{313}\\
u_t&=D_x^2\left[ \frac{u_x}{u^{3/2}}\right], \label{314}\\
u_t&=D_x\left[D_x\left(\frac{u_x}{u^{3/2}}\right)-2\frac{1}{u^{1/2}}\right]\;  \label{315}
\end{align}
\end{linenomath}
respectively.
\end{enumerate}
All these equations are related to the linear equation or to the KdV equation through differential substitutions.
\end{theorem}

\begin{remark}
It is clear that the equations appearing above cannot admit solutions with compact support, let alone compactons. 
As explained in Section 1, this theorem extends and enriches the discussion on $K(m,n)$ appearing in \cite{HHR}. We mention that J. Vodov\'a classified conservation laws of the $K(m,m)$ equations and observed that $K(-2,-2)$, $K(-1/2,-1/2)$ are integrable; integrability of $K(-2,-2)$, $K(-1/2,-1/2)$, and $K(-1,-2)$, is also observed in the later review \cite{RZ}. 
\end{remark}

\section{Integrability of Cooper-Shepard-Sodano}

In this section we study equations of the form 
\begin{equation}
u_{t}+u^{l-2}u_{x}-\alpha pD_{x}\left(u^{p-1}u_{x}^{2}\right)+2\alpha D_{x}^{2}\left(u^{p}u_{x}\right)=0,
                                                                          \qquad\alpha\neq0.\label{eq:eqRHE}
\end{equation}
We consider the case~$p=0$ first. Equation (\ref{eq:eqRHE}) becomes $u_{t}+u^{l-2}u_{x}+2\alpha u_{xxx}=0$,
which is integrable if and only if~$l=2,3,4$, i.e. in the linear, KdV and mKdV case, as observed in \cite{MSS}. 
Let us now consider $p\neq0$. We use the same strategy as with the Rosenau-Hyman case: first we apply the 
change~$t\to -t$, $u\to (2\alpha u^3)^{-1/p}$ to obtain
\begin{equation} \label{css1}
u_{t}
= D_{x}\left[\frac{u_{xx}}{u^3}
-\frac{3 (2 p+3)}{2 p}\frac{u_{x}^2}{u^4}
-\frac{p (2 \alpha )^{\frac{2-l}{p}} }{3 l-p-6}u^{\frac{6-3 l+p}{p}}
\right]+\frac{(p+3) (p+6) u_{x}^3}{2 p^2 u^5}
\end{equation}
if $3l-p-6\neq0$, and
\begin{equation} \label{css2}
u_{t}=D_{x}\left[\frac{u_{xx}}{u^3}
-\frac{3 (2 p+3) }{2 p}\frac{u_{x}^2}{u^4}
+\frac{1}{\sqrt[3]{2 \alpha}}\log{u}\right]
+\frac{(p+3) (p+6) u_{x}^3}{2 p^2 u^5}
\end{equation}
if $3l-p-6=0$.

Thus, the first integrability condition of Proposition \ref{p1} fixes the values~$p=-3$ or~$p=-6$. If~$p=-3$, a combination of potentiation, hodograph,
antipotentiation and exponential point transformation~$u\to{\rm e}^u$ , with a scaling to absorb the constant~$\alpha$, change (\ref{css1}) and (\ref{css2}) into the equations
\[
u_{t}=u_{xxx}-\frac{u_{x}^3}{2}+\frac{(l-2) e^{u-l u} u_{x}}{l-1}
\quad\text{ and }\quad
u_{t}=u_{xxx}-\frac{u_{x}^3}{2}+u u_{x}\; 
\]
respectively.
On the other hand, if~$p=-6$, the same combination of transformations, using~$u\to u^2$ instead of the exponential,
changes (\ref{css1}) and (\ref{css2}) into
\[
u_{t}=u_{xxx}+\frac{(l-2)}{l}\frac{u_{x}}{u^l}\quad\text{ and }\quad
u_{t}=u_{xxx}+\log (u)u_{x}\; 
\]
respectively.

Let us assume that $3l-p-6\neq0$. The integrability condition in~$\rho_1$ implies that if $p=-3$ we must have 
$l=-1,2,3$, and if~$p=-6$, then $l=-2,-1,2$ and we recover again the linear, KdV, mKdV and CDF equations. 
When~$3l-p-6=0$, we have the CSS equations 
\[
u_{t}=\frac{a}{u^{6}}u_{xxx}-12\frac{a}{u^{7}}u_{x}u_{xx}+21\frac{a}{u^{8}}u_{x}^{3}+\frac{u_{x}}{u^{2}}
\]
($p = -6$ and $l=0$) and
\[
u_{t}=\frac{a}{u^{3}}u_{xxx}-6\frac{a}{u^{4}}u_{x}u_{xx}+6\frac{a}{u^{5}}u_{x}^{3}+\frac{u_{x}}{u}\; ,
\]
($p=-3$ and $l=1$), in which $a = 2\alpha$. The integrability condition for $\rho_{2}$ implies that both 
equations are not integrable.

Summarizing, we have the following theorem.
\begin{theorem}
The integrable equations of family \eqref{eq:eqRHE} are
\begin{enumerate}
\item $p=0,$ $l=2,3,4$, corresponding to the linear equation, KdV equation, mKdV equation;
\item $p=-6$, $l=-2,-1,2$, corresponding to the Equations 
\begin{linenomath}
\begin{gather}
u_{t}=\frac{a}{u^{6}}u_{xxx}-12\frac{a}{u^{7}}u_{x}u_{xx}+21\frac{a}{u^{8}}u_{x}^{3}+\frac{u_{x}}{u^{4}},\label{eq:eq6m2}\\
u_{t}=\frac{a}{u^{6}}u_{xxx}-12\frac{a}{u^{7}}u_{x}u_{xx}+21\frac{a}{u^{8}}u_{x}^{3}+\frac{u_{x}}{u^{3}},\label{eq:eq6m1}\\
u_{t}=\frac{a}{u^{6}}u_{xxx}-12\frac{a}{u^{7}}u_{x}u_{xx}+21\frac{a}{u^{8}}u_{x}^{3}+u_{x}.\label{eq:eq6p2}
\end{gather}
\end{linenomath}
respectively.
\item $p=-3,$ $l=-1,2,3$, corresponding to the Equations
\begin{linenomath}
\begin{gather}
u_{t}=\frac{a}{u^{3}}u_{xxx}-6\frac{a}{u^{4}}u_{x}u_{xx}+6\frac{a}{u^{5}}u_{x}^{3}+\frac{u_{x}}{u^{3}},\label{eq:eq3m1}\\
u_{t}=\frac{a}{u^{3}}u_{xxx}-6\frac{a}{u^{4}}u_{x}u_{xx}+6\frac{a}{u^{5}}u_{x}^{3}+u_{x},\label{eq:eq3p2}\\
u_{t}=\frac{a}{u^{3}}u_{xxx}-6\frac{a}{u^{4}}u_{x}u_{xx}+6\frac{a}{u^{5}}u_{x}^{3}+uu_{x}.\label{eq:eq3p3}
\end{gather}
\end{linenomath}
respectively.
\end{enumerate}
All these equations are related to the linear equation or to the KdV equation through differential substitutions.
\end{theorem}

As in Section 3, this theorem implies that no integrable CSS equation admits solutions with compact support.

\section{Integrability and recursion operators}

In this section we construct explicit recursion operators for the equations appearing in the above theorems using
the work \cite{PNN}. First of all, we note that these equations are all in \cite{MSS}. We have (when we write~(4.x.xx) we are referring to the corresponding equation in~\cite{MSS}):
\begin{enumerate}
\item Equation \eqref{eq:eqm2m2} is a special case of (4.1.27), namely,
\begin{equation} \label{eq:eqm2m21}
u_t=D_x\left(\frac{u_{xx}}{u^3}-3\frac{u_x^2}{u^4}+\frac{1}{2u^2}\right) =-9\,{\frac {u_{{x}}u_{{xx}}}{{u}^{4}}}
+12\,{\frac {{u_{{x}}}^{3}}{{u}^{5}}}-{\frac {u_{{x}}}{{u}^{3}}}+{\frac {u_{{xxx}}}{{u}^{3}}}
\; .
\end{equation}
\item Equation (\ref{39}) is equivalent to a subcase of~(4.1.25)
\begin{equation} \label{391}
u_t=D_x\!\left(\frac{u_{xx}}{u^3}{-}3\frac{u_x^2}{u^4}{-}\frac{3}{2u}{+}cu\right)= -9\,{\frac {u_{{x}}u_{{xx}}}{{u}^{4}}}
+12\,{\frac {{u_{{x}}}^{3}}{{u}^{5}}}+{\frac {3 u_{{x}}}{2 {u}^{2}}}+ c u_x + {\frac {u_{{xxx}}}{{u}^{3}}}\; .
\end{equation}
\item Equations (\ref{395}) and (\ref{310}) are equivalent to subcases of~(4.1.34)
\begin{linenomath}
\begin{multline} \label{3951}
u_t=D_x\left(\frac{u_{xx}}{u^3}-3\frac{u_x^2}{u^4}+c_1\frac{u_x}{u^2}+c_2u\right) 
\\
= 
-9\,{\frac {u_{{x}}u_{{xx}}}{{u}^{4}}}
+12\,{\frac {{u_{{x}}}^{3}}{{u}^{5}}}-2c_1{\frac {u_{{x}}^2}{{u}^{3}}}+ c_2 u_x +c_1\frac{u_{xx}}{u^2} 
+ {\frac {u_{{xxx}}}{{u}^{3}}} \; .
\end{multline}
\end{linenomath}
\item Equations (\ref{312})-(\ref{315}) are all equivalent to subcases of~(4.1.30)
\[
u_t=D_x\!\left[\frac{u_{xx}}{u^3}-\frac32\frac{u_x^2}{u^4}-\frac32\frac{\lambda u_x^2}{u^3(\lambda u+1)}+
                                c_1\frac{(\lambda u+1)^3}{u^2}+c_2\frac{u^2}{\lambda u+1}+c_3u\right]
\]
with~$\lambda=0$, that is,
\begin{equation} \label{2171}
u_t=-6\,{\frac {u_{x}u_{{xx}}}{{u}^{4}}}
+6\,{\frac{{u_{{x}}}^{3}}{{u}^{5}}}
-2 {\it c_1}\,{\frac {u_{{x}}}{{u}^{3}}}+2{\it c_2}\,u u_{x}
+{\it c_3} u_{{x}}+{\frac {u_{{xxx}}}{{u}^{3}}} \; ,
\end{equation}
after applying the point transformation~$u\to u^2$.
\end{enumerate}

Now, Equations \eqref{eq:eqm2m21}, \eqref{391} and \eqref{3951} are special cases (for the values of the
constant numbers $c_1,c_2$ relevant to us) of Equation (81) in \cite{PNN}, namely,
\begin{equation} \label{class-0}
u_t = \frac{u_{xxx}}{u^3} - 9 \frac{u_x u_{xx}}{u^4} + 12 \frac{u_x^3}{u^5} 
+ \frac{2}{3}\lambda_1 \frac{u_x}{u^3} + \frac{1}{2} \lambda_2 \frac{u_x}{u^2} - c u_x \; ,
\end{equation} 
while Equation (\ref{2171}) is Equation (85) in \cite{PNN}, see below. The recursion operator for Equation~(\ref{class-0})
is as follows:
\begin{linenomath}
\begin{multline*}
  R[u]=u^{-2}D_x^2
  -5u^{-3}u_xD_x
  -4u^{-3}u_{xx}+12u^{-4}u_x^2
  +\frac{2\lambda_1}{3}u^{-2}
  +\frac{2\lambda_2}{3}u^{-1}\\
  -u_tD_x^{-1}\circ 1
  +u_xD_x^{-1}\circ \left(\frac{\lambda_2}{6}u^{-2}-c\right).
\end{multline*}
\end{linenomath}
Acting $R[u]$ on the $t$-translation symmetry $\displaystyle{u_t\pde{\ }{u}}$, yields a corresponding 
symmetry-integrable hierarchy of order $2m+3$, namely
\begin{linenomath}
\begin{gather*}
  u_{t}=R^m[u]\circ  \left(  \frac{u_{xxx}}{u^3} - 9 \frac{u_x u_{xx}}{u^4} + 12 \frac{u_x^3}{u^5} 
+ \frac{2}{3}\lambda_1 \frac{u_x}{u^3} + \frac{1}{2} \lambda_2 \frac{u_x}{u^2} - c u_x  \right),\\[2mm]
  m=0,1,2,\ldots \; ,
\end{gather*}
\end{linenomath}
and we note that for the $x$-translation symmetry we obtain
\[  R[u]\circ u_x=0. \]
\ \\
\indent Let us now consider the integrable cases of the CSS equations. We see that Equations $(\ref{eq:eq6m2})$, 
$(\ref{eq:eq6m1})$ and $(\ref{eq:eq6p2})$  are special cases of Equation (90) in \cite{PNN}, namely
\begin{linenomath}
\begin{equation}
  \label{class-1}
  u_t=\frac{au_{xxx}}{u^6}-12\frac{au_xu_{xx}}{u^7}+21\frac{au_x^3}{u^8}
  +\beta_1 \frac{u_x}{u^4}+\beta_2 \frac{u_x}{u^3}+\beta_3 u_x,
\end{equation}  
\end{linenomath}
where $a$, $\beta_1$, $\beta_2$ and $\beta_3$ are arbitrary constants and $a\neq 0$. This equation admits the following 
recursion operator:
\begin{linenomath}
\begin{multline} 
  R_1[u]=\frac{1}{u^4}D_x^2
  -\frac{6u_x}{u^5}D_x
  -\frac{6u_{xx}}{u^5}
  +\frac{22u_x^2}{u^6}
  +\frac{2\beta_2}{a}   \frac{1}{u}
  +\frac{4\beta_1}{3a}  \frac{1}{u^2}
  -\frac{2}{a}u_tD_x^{-1}\circ u 
  \\
 +u_xD_x^{-1}\circ
  \left(
  -\frac{2u_{xx}}{u^6}
  +6\frac{u_x^2}{u^7}
  +\frac{\beta_2}{a}  \frac{1}{u^2}
  +\frac{2\beta_1}{3a} \frac{1}{u^3}
  +\frac{2\beta_3}{a}u\right).\label{R1}
\end{multline}
\end{linenomath}
Acting $R_1[u]$ on the $t$-translation symmetry $\displaystyle{u_t\pde{\ }{u}}$, we obtain a corresponding 
symmetry-integrable hierarchy of order $2m+3$, namely
\begin{linenomath}
\begin{gather}
  \label{hierarchy-1}
  u_{t}=R_1^m[u]\circ  \left(\frac{au_{xxx}}{u^6}-12\frac{au_xu_{xx}}{u^7}+21\frac{au_x^3}{u^8}
  +\beta_1 \frac{u_x}{u^4}+\beta_2 \frac{u_x}{u^3}+\beta_3 u_x\right),\\[2mm]
  m=0,1,2,\ldots \nn \; ,
\end{gather}
\end{linenomath}
and we note that for the $x$-translation symmetry we obtain
\[
  R_1[u]\circ u_x=0.
\]
\ \\
\indent On the other hand, Equations \eqref{eq:eq3m1}, \eqref{eq:eq3p2} and \eqref{eq:eq3p3} are special cases of 
Equation~(85) in \cite{PNN}, namely
\begin{equation}
\label{class-2}
  u_t=\frac{au_{xxx}}{u^3}
  -6\frac{au_xu_{xx}}{u^4}
  +6\frac{au_x^3} {u^5}
    +\beta_1\frac{u_x}{u^3}
    +\beta_2 uu_x
    +\beta_3u_x,
\end{equation}
where $a$, $\beta_1$, $\beta_2$ and $\beta_3$ are arbitrary constants and $a\neq 0$. This equation admits the following 
recursion operator:
\begin{linenomath}
\begin{multline}\label{R2}
  R_2[u]=\frac{1}{u^2}D_x^2
  -\frac{3u_x}{u^3} D_x
  -3\frac{u_{xx}}{u^3}
  +6\frac{u_x^2}{u^4}
  +\frac{\beta_2}{3a}  u^2
  +\frac{\beta_1}{a} \frac{1}{u^2}\\[0.2cm]
  -\frac{1}{a} u_t\,D_x^{-1}\circ 1
  +\frac{4\beta_2}{3a} u_x\,D_x^{-1}\circ u
  +\frac{\beta_3}{a} u_x\,D_x^{-1}\circ 1.
\end{multline}
\end{linenomath}
Acting $R_2[u]$ on the $t$-translation symmetry $\displaystyle{u_t\pde{\ }{u}}$, we obtain a symmetry-integrable hierarchy of order $2m+3$, namely 
\begin{linenomath}
\begin{gather}
  \label{hierarchy-2}
  u_{t}=R_2^m[u]\circ  \left(
\frac{au_{xxx}}{u^3}
  -6\frac{au_xu_{xx}}{u^4}
  +6\frac{au_x^3} {u^5}
    +\beta_1\frac{u_x}{u^3}
    +\beta_2 uu_x
    +\beta_3u_x
    \right),\\[2mm]
 m=0,1,2,\ldots\nn \; ,
 \end{gather}
\end{linenomath}
and we note that for the $x$-translation symmetry we obtain
\[  R_2[u]\circ u_x=0.
\]

\section{The isochronous equations for (\ref{class-1}) and (\ref{class-2})}
In this section we construct new integrable evolution equations starting from what we will call the Cooper-Shepard-Sodano model equations
(\ref{class-1}) and (\ref{class-2}). Our new equations are isochronous in the sense of Calogero, see \cite[Chapter 7]{Calogero-2005} and 
\cite{Mariani-Calogero-2005, Calogero-Mariani-2005, Calogero-Euler-Euler}: they are autonomous evolution PDEs which depend on a positive 
parameter $\omega$ and possess {\em many} solutions which are time-periodic with period $T=2\pi/\omega$. For completeness, we also explain how to obtain the Lie point symmetries of our equations and present their recursion operators.

\subsection{Isochronous equations}
Following \cite{Mariani-Calogero-2005,Calogero-Mariani-2005,Calogero-2005,Calogero-Euler-Euler}, we introduce
a new dependent variable $v(r,s)$, where $r$ and $s$ are new independent variables, as follows:
\begin{linenomath}
\begin{subequations}
  \begin{gather}
    \label{Iso-transformation-u}
u(x,t)=e^{-i\lambda \omega s} v(r,s)\\[0.2cm]
\label{Iso-transformation-x}
x=re^{i\mu \omega s}\\[0.2cm]
\label{Iso-transformation-t}
t=\frac{1}{i\omega}\left(e^{i\omega s}-1\right).
\end{gather}
\end{subequations}
\end{linenomath}
The prolongations are
\begin{linenomath}
\begin{subequations}
\begin{gather*}
  u_t=e^{-i(\lambda+1)\omega s}\left[v_s-i\lambda \omega\, v-i\mu \omega r\, v_r\right]\\[0.2cm]
  u_{nx}=e^{-i(\lambda+n\mu)\omega s}\, v_{nr},\quad n=1,2,3,\dots,
\end{gather*}
\end{subequations}
\end{linenomath}
where
\[
  u_{nx}=\frac{\p^n u}{\p x^n},\qquad   v_{nr}=\frac{\p^n v}{\p r^n}.
\]
With the change of variables \eqref{Iso-transformation-u} -- \eqref{Iso-transformation-t}, equation \eqref{class-1}
takes the form
\begin{linenomath}
\begin{multline}
  v_s-i\lambda \omega v-i\mu\omega r v_r=
  e^{i(6\lambda-3\mu+1)\omega s}
  \left(\frac{av_{rrr}}{v^6}-\frac{12av_rv_{rr}}{v^7}+\frac{21av_r^3}{v^8}\right)\\[0.2cm]
  \label{Iso-class-1}
\qquad  +\beta_1e^{i(4\lambda -\mu+1)\omega s}\,\left(\frac{v_r}{v^4}\right)
  +\beta_2e^{i(3\lambda-\mu+1)\omega s}\,\left(\frac{v_r}{v^3}\right)
  +\beta_3 e^{i(-\mu+1)\omega s}\,v_r.
\end{multline}
\end{linenomath}
This equation can become autonomous for $a\neq 0$ only if
\[
  \mu=2\lambda+\frac{1}{3},
\]
so that (\ref{Iso-class-1}) then takes the form
\begin{linenomath}
\begin{multline}
  v_s-i\lambda \omega v-i\left(2\lambda+\frac{1}{3}\right)  \omega r v_r=
  \frac{av_{rrr}}{v^6}-\frac{12av_rv_{rr}}{v^7}+\frac{21av_r^3}{v^8}\\[0.2cm]
  \label{Iso-subclass-1}
  +\beta_1e^{i(2\lambda+\frac{2}{3})\omega s}\,\left(\frac{v_r}{v^4}\right)
  +\beta_2e^{i(\lambda+\frac{2}{3})\omega s}\,\left(\frac{v_r}{v^3}\right)
  +\beta_3 e^{i(2\lambda+\frac{2}{3})\omega s}\,v_r.
\end{multline}
\end{linenomath}
Clearly (\ref{Iso-subclass-1}), and therefore (\ref{Iso-class-1}), becomes autonomous in the following three cases:
\ \\[2mm]
\noindent
    {\bf Case 1.1:} $\displaystyle{\lambda=-\frac{1}{3}}$ and $\displaystyle{\mu=-\frac{1}{3}}$ with $\beta_2=\beta_3=0$.
    Then (\ref{Iso-class-1}) becomes
    \begin{equation}
      \label{Iso-class-1.1}
  v_s+i\frac{1}{3}\omega v+i\frac{1}{3} \omega r v_r=
  \frac{av_{rrr}}{v^6}-\frac{12av_rv_{rr}}{v^7}+\frac{21av_r^3}{v^8}
  +\beta_1 \frac{v_r}{v^4}.
\end{equation}       
 {\bf Case 1.2:} $\displaystyle{\lambda=-\frac{2}{3}}$ and $\displaystyle{\mu=-1}$ with $\beta_1=\beta_3=0$.
 Then (\ref{Iso-class-1}) becomes
 \begin{equation}
    \label{Iso-class-1.2}
  v_s+i\frac{2}{3} \omega v+i \omega r v_r=
  \frac{av_{rrr}}{v^6}-\frac{12av_rv_{rr}}{v^7}+\frac{21av_r^3}{v^8}
   +\beta_2\frac{v_r}{v^3}.
\end{equation}  
{\bf Case 1.3:} $\displaystyle{\lambda=\frac{1}{3}}$ and $\displaystyle{\mu=1}$ with $\beta_1=\beta_2=0$.
 Then (\ref{Iso-class-1}) becomes
 \begin{equation}
    \label{Iso-class-1.3}
  v_s-i\frac{1}{3} \omega v-i \omega r v_r=
  \frac{av_{rrr}}{v^6}-\frac{12av_rv_{rr}}{v^7}+\frac{21av_r^3}{v^8}+\beta_3 v_r.
\end{equation}  
\ \\
\indent
Now we consider (\ref{class-2}). With the change of variables (\ref{Iso-transformation-u}) -- (\ref{Iso-transformation-t}), Equation~\eqref{class-2}
takes the form
\begin{linenomath}
\begin{multline}
  v_s-i\lambda \omega v-i\mu\omega r v_r=
  e^{i(3\lambda-3\mu+1)\omega s}
  \left(\frac{av_{rrr}}{v^3}-\frac{6av_rv_{rr}}{v^4}+\frac{6av_r^3}{v^5}\right)\\[0.2cm]
  \label{Iso-class-2}
\qquad  +\beta_1e^{i(3\lambda -\mu+1)\omega s}\,\left(\frac{v_r}{v^3}\right)
  +\beta_2e^{i(-\lambda-\mu+1)\omega s}\,vv_r
  +\beta_3 e^{i(-\mu+1)\omega s}\,v_r.
\end{multline}
\end{linenomath}
This equation can become autonomous for $a\neq 0$ only if
\[
  \mu=\lambda+\frac{1}{3},
\]
so that (\ref{Iso-class-2}) then takes the form
\begin{linenomath}
\begin{multline}
  v_s-i\lambda \omega v-i\left(\lambda+\frac{1}{3}\right)  \omega r v_r=
  \frac{av_{rrr}}{v^3}-\frac{6av_rv_{rr}}{v^4}+\frac{6av_r^3}{v^5}\\[0.2cm]
  \label{Iso-subclass-2}
\qquad  +\beta_1e^{i(2\lambda+\frac{2}{3})\omega s}\,\left(\frac{v_r}{v^3}\right)
  +\beta_2e^{i(-2\lambda+\frac{2}{3})\omega s}\,vv_r
  +\beta_3 e^{i(-\lambda+\frac{2}{3})\omega s}\,v_r.
\end{multline}  
\end{linenomath}
Clearly (\ref{Iso-subclass-2}), and therefore (\ref{Iso-class-2}), becomes autonomous in the following three cases:\\

\noindent
 {\bf Case 2.1:} $\displaystyle{\lambda=-\frac{1}{3}}$ and $\displaystyle{\mu=0}$ with $\beta_2=\beta_3=0$.
Then (\ref{Iso-class-2}) becomes
\begin{equation}
      \label{Iso-class-2.1}
    v_s+i\frac{1}{3} \omega v=
  \frac{av_{rrr}}{v^3}-\frac{6av_rv_{rr}}{v^4}+\frac{6av_r^3}{v^5}
  +\beta_1\,\frac{v_r}{v^3}.
\end{equation}       
 {\bf Case 2.2:} $\displaystyle{\lambda=\frac{1}{3}}$ and $\displaystyle{\mu=\frac{2}{3}}$ with $\beta_1=\beta_3=0$.
 Then (\ref{Iso-class-2}) becomes
 \begin{equation}
   \label{Iso-class-2.2}
 v_s-i\frac{1}{3} \omega v-i\frac{2}{3}  \omega r v_r=
  \frac{av_{rrr}}{v^3}-\frac{6av_rv_{rr}}{v^4}+\frac{6av_r^3}{v^5}
  +\beta_2\,vv_r.
\end{equation}  
{\bf Case 2.3:} $\displaystyle{\lambda=\frac{2}{3}}$ and $\displaystyle{\mu=1}$ with $\beta_1=\beta_2=0$.
 Then (\ref{Iso-class-2}) becomes
 \begin{equation}
   \label{Iso-class-2.3}
  v_s-i\frac{2}{3} \omega v-i\omega r v_r=
  \frac{av_{rrr}}{v^3}-\frac{6av_rv_{rr}}{v^4}+\frac{6av_r^3}{v^5}
  +\beta_3 \,v_r.  
\end{equation}  

Our isochronous equations are (\ref{Iso-class-1.1})--(\ref{Iso-class-1.3}) and (\ref{Iso-class-2.1})--(\ref{Iso-class-2.3}).

\subsection{Symmetries}
We list the Lie point symmetries of equation (\ref{class-1}), that is
\[
  u_t=\frac{au_{xxx}}{u^6}-12\frac{au_xu_{xx}}{u^7}+21\frac{au_x^3}{u^8}
  +\beta_1 \frac{u_x}{u^4}+\beta_2 \frac{u_x}{u^3}+\beta_3 u_x.
\]
Besides the obvious $x$-translation, $\displaystyle{Z_x=\pde{\ }{x}}$, and $t$-translation symmetry,
$\displaystyle{Z_t=\pde{\ }{t}}$,
 Equation~\eqref{class-1} also admits the following point symmetries:
\begin{itemize}
\item[a)] For $\beta_1=\beta_2=\beta_3=0$:
\[
    Z_1=x\pde{\ }{x}+3t\pde{\ }{t},\quad Z_2=u\pde{\ }{u}+6t\pde{\ }{t}.
\]

\item[b)] For $\beta_1=\beta_2=0$ and $\beta_3\neq 0$:
\[
    Z_1=(x-2t\beta_3)\pde{\ }{x}+3t\pde{\ }{t},\quad Z_2=u\pde{\ }{u}+6t\pde{\ }{t}-6t\beta_3\pde{\ }{x}.
\]

\item[c)] For $\beta_1=\beta_3=0$ and $\beta_2\neq 0$:
\[
    Z_1=-\frac{2}{3}u\pde{\ }{u}+x\pde{\ }{x}-t\pde{\ }{t}.
\]
  
\item[d)] For $\beta_2=\beta_3=0$ and $\beta_1\neq 0$:
\[
    Z_1=-u\pde{\ }{u}+x\pde{\ }{x}-3t\pde{\ }{t}.
\]
\end{itemize}  
\ \\
\indent We list the Lie point symmetries of equation (\ref{class-2}), that is
\[
  u_t=\frac{au_{xxx}}{u^3}
  -6\frac{au_xu_{xx}}{u^4}
  +6\frac{au_x^3} {u^5}
    +\beta_1\frac{u_x}{u^3}
    +\beta_2 uu_x
    +\beta_3u_x.
\]
Besides the obvious $x$-translation symmetry and $t$-translation symmetry,
(\ref{class-2}) also admits the following point symmetries:
\begin{itemize}

\item[a)] For $\beta_1=\beta_2=\beta_3=0$:
\[
    Z_1=x\pde{\ }{x}+3t\pde{\ }{t},\quad Z_2=u\pde{\ }{u}+3t\pde{\ }{t},\quad
    Z_3=xu\pde{\ }{u}-\frac{1}{2}x^2\pde{\ }{x}.
\]

\item[b)] For $\beta_1=\beta_2=0$ and $\beta_3\neq 0$:
\begin{linenomath}
\begin{gather*}
    Z_1=(x-2t\beta_3)\pde{\ }{x}+3t\pde{\ }{t},\quad Z_2=u\pde{\ }{u}+3t\pde{\ }{t}-3t\beta_3\pde{\ }{x}\\[2mm]
    Z_3=u(x+\beta_3t)\pde{\ }{u}-\frac{1}{2}(x+\beta_3t)^2\pde{\ }{x}.
\end{gather*}
\end{linenomath}

  \item[c)] For $\beta_1=\beta_3=0$ and $\beta_2\neq 0$:
\[
    Z_1=u\pde{\ }{u}-2x\pde{\ }{x}-3t\pde{\ }{t}.
\]

\item[d)] For $\beta_2=\beta_3=0$ and $\beta_1\neq 0$:
\begin{linenomath}
    \begin{gather*}
      Z_1=u\pde{\ }{u}+3t\pde{\ }{t}\\[2mm]
      Z_2=u\sin\left(a^{-1/2}\beta_1^{1/2}  x\right)\pde{\ }{u}+a^{1/2}\beta_1^{-1/2}\cos\left(a^{-1/2}\beta_1^{1/2} x\right)\pde{\ }{x}\\[2mm]
Z_3=u\cos\left(a^{-1/2}\beta_1^{1/2}x\right)\pde{\ }{u}-a^{1/2}\beta_1^{-1/2}\sin\left(a^{-1/2}\beta_1^{1/2}x\right)\pde{\ }{x}.
  \end{gather*}
\end{linenomath}
  
\end{itemize}

We can now obviously map the symmetries of (\ref{class-1}) and (\ref{class-2}) with the
 (\ref{Iso-transformation-u}) -- (\ref{Iso-transformation-t})
to symmetries of the isochronous equations (\ref{Iso-class-1}) and (\ref{Iso-class-2}).
For example, in vertical form the $x$-translation symmetry 
\[
   u_x\pde{\ }{u},
\]
 then takes the form
\[
   e^{-i\mu \omega s}v_r\pde{\ }{v},
\]
 for (\ref{Iso-class-1}) and (\ref{Iso-class-2}),
 whereas the $t$-translation symmetry
\[
   u_t\pde{\ }{u},
\]
becomes the symmetry
\[
  e^{-i\omega s}\left(v_s-i\lambda \omega v-i\mu\omega rv_r\right)\pde{\ }{v}
\]
for (\ref{Iso-class-1}) and (\ref{Iso-class-2}).

\subsection{The isochronous hierarchies for (\ref{class-1}) and (\ref{class-2})}

For equation (\ref{class-1}) we have the hierarchy (\ref{hierarchy-1}), namely
\begin{linenomath}
\begin{gather*}
u_{t}=R_1^m[u]\circ 
\left(\frac{au_{xxx}}{u^6}-12\frac{au_xu_{xx}}{u^7}+21\frac{au_x^3}{u^8}
+\beta_1 \frac{u_x}{u^4}+\beta_2 \frac{u_x}{u^3}+\beta_3 u_x,\right)\\[2mm]
m=0,1,2,\ldots\ ,\nonumber  
\end{gather*}
\end{linenomath}
where $R_1[u]$ is given by \eqref{R1}.
Corresponding to the above Case 1.1, Case 1.2 and Case 1.3, the isochronous hierarchies are the following:\\
\ \\
\noindent
    {\bf Case 1.1:} We consider the hierarchy (\ref{hierarchy-1}) with $\beta_2=\beta_3=0$. This leads to the following isochronous hierarchy
\begin{linenomath}
    \begin{multline}
  v_{s}+i\left(\frac{1}{2m+3}\right) \omega v+i\left(\frac{1}{2m+3}\right)\omega rv_r\\
\label{Iso-hierarchy-1.1}  
  \qquad
=R^m_{11}[v]\circ
  \left( \frac{av_{rrr}}{v^6}-\frac{12av_rv_{rr}}{v^7}+\frac{21av_r^3}{v^8}
  +\beta_1 \frac{v_r}{v^4} \right),\quad m=0,1,2,\ldots\ ,
\end{multline}
\end{linenomath}  
where
\begin{linenomath}
\begin{multline*}
  R_{11}[v]=
 \frac{1}{v^4}D_r^2
  -\frac{6v_r}{v^5}D_r
  -\frac{6v_{rr}}{v^5}
  +\frac{22v_r^2}{v^6}
  +\frac{4\beta_1}{3a}  \frac{1}{v^2}\\
\qquad   -\frac{2}{a}
  \left(
\frac{av_{rrr}}{v^6}-\frac{12av_rv_{rr}}{v^7}+\frac{21av_r^3}{v^8}
  +\beta_1 \frac{v_r}{v^4}
  \right)
  D_r^{-1}\circ v \\
\qquad 
 +v_rD_r^{-1}\circ
  \left(
  -\frac{2v_{rr}}{v^6}
  +6\frac{v_r^2}{v^7}
  +\frac{2\beta_1}{3a} \frac{1}{v^3} \right).
\end{multline*}
\end{linenomath}
The first member of the hierarchy~\eqref{Iso-hierarchy-1.1} for $m=0$ is the equation (\ref{Iso-class-1.1}).\\     
\ \\
\noindent
    {\bf Case 1.2:} We consider the hierarchy (\ref{hierarchy-1}) with $\beta_1=\beta_3=0$. This leads to the following isochronous hierarchy
\begin{linenomath}
    \begin{multline}
  v_{s}+i\left(\frac{2}{2m+3}\right) \omega v+i\left(\frac{3}{2m+3}\right)\omega rv_r\\
\label{Iso-hierarchy-1.2}  
  \qquad
=R^m_{12}[v]\circ
  \left( \frac{av_{rrr}}{v^6}-\frac{12av_rv_{rr}}{v^7}+\frac{21av_r^3}{v^8}
  +\beta_1 \frac{v_r}{v^4} \right),\quad m=0,1,2,\ldots\ ,
\end{multline}
\end{linenomath}  
where
\begin{linenomath}
\begin{multline*}
  R_{12}[v]=\frac{1}{v^4}D_r^2
  -\frac{6v_r}{v^5}D_r
  -\frac{6v_{rr}}{v^5}
  +\frac{22v_r^2}{v^6}
  +\frac{2\beta_2}{a}   \frac{1}{v}\\
  \qquad
  -\frac{2}{a}
\left(
\frac{av_{rrr}}{v^6}
-\frac{12av_rv_{rr}}{v^7}
+\frac{21av_r^3}{v^8}
+\beta_2 \frac{v_r}{v^3}
\right)
  D_r^{-1}\circ v \\
  \qquad 
 +v_rD_r^{-1}\circ
  \left(
  -\frac{2v_{rr}}{v^6}
  +6\frac{v_r^2}{v^7}
  +\frac{\beta_2}{a}  \frac{1}{v^2}
  \right).
\end{multline*}
\end{linenomath}
The first member of the hierarchy~\eqref{Iso-hierarchy-1.2} for $m=0$ is the equation (\ref{Iso-class-1.2}).\\
\ \\ 
\noindent
    {\bf Case 1.3:} We consider the hierarchy (\ref{hierarchy-1}) with $\beta_1=\beta_2=0$. This leads to the following isochronous hierarchy
\begin{linenomath}
\begin{multline}
  v_{s}+i\lambda \omega v+i\left(2\lambda+\frac{1}{2m+3} \right)\omega rv_r\\
\label{Iso-hierarchy-1.3}  
  \qquad
=R^m_{13}[v]\circ
  \left( \frac{av_{rrr}}{v^6}-\frac{12av_rv_{rr}}{v^7}+\frac{21av_r^3}{v^8}
  +\beta_3 v_r \right),\quad m=1,2,\ldots
\end{multline}
\end{linenomath}  
where $\lambda$ is arbitrary and
\begin{linenomath}
\begin{multline*}
  R_{13}[v]=\frac{1}{v^4}D_r^2
  -\frac{6v_r}{v^5}D_r
  -\frac{6v_{rr}}{v^5}
  +\frac{22v_r^2}{v^6}\\
  \qquad
  -\frac{2}{a}
\left(
\frac{av_{rrr}}{v^6}
-\frac{12av_rv_{rr}}{v^7}
+\frac{21av_r^3}{v^8}
+\beta_3 v_r
\right)
  D_r^{-1}\circ v \\
\label{R13}
  \qquad 
 +v_rD_r^{-1}\circ
  \left(
  -\frac{2v_{rr}}{v^6}
  +6\frac{v_r^2}{v^7}
  +\frac{2\beta_3}{a}v_r\right).
\end{multline*}
\end{linenomath}
Note that equation (\ref{Iso-class-1.3}) does {\bf not} correspond to $m=0$ in \eqref{Iso-hierarchy-1.3}.
The reason is rather obvious: since
\[  R^m_{13}[v] \circ \left(\beta_3 v_r\right)=0
\]
for all  $m=1,2,\ldots$, the $\beta_3$ term disappears in (\ref{Iso-hierarchy-1.3}) and there remains only one constraint on $\lambda$ and $\mu$ to assure
that the hierarchy does not depend explicitly on $s$,  namely 
\[
  \mu-2\lambda-\frac{1}{2m+3}=0.
\]
\ \\
\indent For the equation (\ref{class-2}) we have the hierarchy (\ref{hierarchy-2}), namely
\begin{linenomath}
\begin{gather*}
  u_{t}=R_2^m[u]\circ  \left(
\frac{au_{xxx}}{u^3}
  -6\frac{au_xu_{xx}}{u^4}
  +6\frac{au_x^3} {u^5}
    +\beta_1\frac{u_x}{u^3}
    +\beta_2 uu_x
    +\beta_3u_x
    \right),\\[2mm]
m=0,1,2,\ldots\ ,\nn
\end{gather*}
\end{linenomath}
where $R_2[u]$ is given by \eqref{R2}.
Corresponding to the above Case 2.1, Case 2.2 and Case 2.3, the isochronous hierarchies are the following:\\
\ \\
\noindent
    {\bf Case 2.1:} We consider the hierarchy (\ref{hierarchy-2}) with $\beta_2=\beta_3=0$. This leads to the following
    isochronous hierarchy
\begin{linenomath}
\begin{gather}
\label{Iso-hierarchy-2.1}
v_{s}+i\left(\frac{1}{2m+3}\right) \omega v  =
R_{21}^m[v]\circ \left(  \frac{av_{rrr}}{v^3}-\frac{6av_rv_{rr}}{v^4}+\frac{6av_r^3}{v^5}
+\beta_1\,\frac{v_r}{v^3}  \right)\\[2mm]
m=0,1,2,\ldots,\nn
\end{gather}
\end{linenomath}
where 
\begin{linenomath}
\begin{multline*}
  R_{21}[v]=\frac{1}{v^2}D_r^2
  -\frac{3v_r}{v^3} D_r
  -3\frac{v_{rr}}{v^3}
  +6\frac{v_r^2}{v^4}
  +\frac{\beta_1}{a} \frac{1}{v^2}\\
  \qquad
  -\frac{1}{a}
\left(  
  \frac{av_{rrr}}{v^3}-\frac{6av_rv_{rr}}{v^4}+\frac{6av_r^3}{v^5}
  +\beta_1\,\frac{v_r}{v^3} 
\right)
  \,D_r^{-1}\circ 1.
\end{multline*}
\end{linenomath}
\ \\
\noindent
    {\bf Case 2.2:} We consider the hierarchy (\ref{hierarchy-2}) with $\beta_1=\beta_3=0$.
    This leads to the following isochronous hierarchy
\begin{linenomath}
\begin{gather}
\begin{split}
v_{s}-i\left(\frac{1}{2m+3}\right) \omega v&-i\left(\frac{2}{2m+3}\right)\omega r v_r 
=
\\
&=R_{22}^m[v]\circ \left(
\frac{av_{rrr}}{v^3}-\frac{6av_rv_{rr}}{v^4}+\frac{6av_r^3}{v^5}+\beta_2 vv_r\right)
\end{split}\label{Iso-hierarchy-2.2}\\[2mm]
m=0,1,2,\ldots,\nn
\end{gather}
\end{linenomath}
where
\begin{linenomath}
\begin{multline}
\label{R22}  
  R_{22}[v]=\frac{1}{v^2}D_r^2
  -\frac{3v_r}{v^3} D_r
  -3\frac{v_{rr}}{v^3}
  +6\frac{v_r^2}{v^4}
 +\frac{\beta_2}{3a}  v^2\\
  \qquad
  -\frac{1}{a}
\left(  
\frac{av_{rrr}}{v^3}-\frac{6av_rv_{rr}}{v^4}+\frac{6av_r^3}{v^5}
+\beta_2 vv_r
\right)
  \,D_r^{-1}\circ 1
  +\frac{4\beta_2}{3a} v_r\,D_r^{-1}\circ v.
\end{multline}
\end{linenomath}
\ \\
\noindent
    {\bf Case 2.3:} We consider the hierarchy (\ref{hierarchy-2}) with $\beta_1=\beta_2=0$.
    This leads to the following isochronous hierarchy
\begin{linenomath}
\begin{gather}
\label{Iso-hierarchy-2.3}
\begin{split}
v_{s}-i\lambda \omega v&-i\left(\lambda +\frac{1}{2m+3}\right)\omega r v_r \\
&\qquad\qquad=R_{23}^m[v]\circ \left(
\frac{av_{rrr}}{v^3}-\frac{6av_rv_{rr}}{v^4}+\frac{6av_r^3}{v^5}+\beta_3 v_r\right)
\end{split}\\[2mm]
m=1,2,3,\ldots,\nn
\end{gather}
\end{linenomath}
where 
\begin{linenomath}
\begin{multline*}
  R_{23}[v]=\frac{1}{v^2}D_r^2
  -\frac{3v_r}{v^3} D_r
  -3\frac{v_{rr}}{v^3}
  +6\frac{v_r^2}{v^4}\\
  \qquad
  -\frac{1}{a}
\left(  
\frac{av_{rrr}}{v^3}-\frac{6av_rv_{rr}}{v^4}+\frac{6av_r^3}{v^5}
+\beta_3 v_r
\right)
  \,D_r^{-1}\circ 1
  +\frac{\beta_3}{a} v_r\,D_r^{-1}\circ 1.
\end{multline*}
\end{linenomath}
Note that equation (\ref{Iso-class-2.3}) does {\bf not} correspond to $m=0$ in (\ref{Iso-hierarchy-2.3}) for the same reason as in
Case 1.3. That is, since
\[
  R^m_{23}[v] \circ \left(\beta_3 v_r\right)=0
\]
for all  $m=1,2,\ldots$, the $\beta_3$ term disappears in (\ref{Iso-hierarchy-2.3}) and there remains only one
constraint on $\lambda$ and $\mu$ to assure
that the hierarchy does not depend explicitly on $s$,  namely 
\[
  \mu-2\lambda-\frac{1}{2m+3}=0.
\]
\ \\
\noindent
{\bf Acknowledgements:} E.G.R. has been partially supported by the FONDECYT operating grant \# 1161691. R.H.H. has been partially supported by the ``Ministerio de Econom\'ia y Competitividad" (MINECO, Spain) under grant MTM2016-79639-P (AEI/FEDER, EU).

\let\thefootnote\relax\footnotetext{\emph{E-mail address:}\/ rafahh@etsist.upm.es}
\let\thefootnote\relax\footnotetext{\emph{E-mail address:}\/ marianna@ltu.se}
\let\thefootnote\relax\footnotetext{\emph{E-mail address:}\/ norbert@ltu.se}
\let\thefootnote\relax\footnotetext{\emph{E-mail address:}\/ enrique.reyes@usach.cl, e\_g\_reyes@yahoo.ca}


\begin{thebibliography}{00}
\bibitem{MAdler} M. Adler, On the trace functional for formal pseudodifferential operators and the
                 symplectic structure of the KdV type equations, {\em Inventiones Math.} 50 (1979),
                 219--248.
                 
\bibitem{AW} D.M. Ambrose and J.D. Wright, Preservation of support and positivity for solutions of
             degenerate evolution equations. {\em Nonlinearity} 23 (2010) 607--620.
             
\bibitem{AW2} D.M. Ambrose and J.D. Wright, Dispersion versus anti-diffusion:well-posedness in variable coefficient and quasilinear equations of KdV type. {\em Indiana Univ. Math. J.} 62 (2013), 1237--1281.
             
\bibitem{ASWY} D.M. Ambrose, G. Simpson, J.D. Wright and D.G Yang, Ill-posedness of degenerate dispersive equations.
               {\em Nonlinearity} 25 (2012) 2655--2680.
               
\bibitem{Bakirov} I.M. Bakirov, On the symmetries of some system of evolution equations.
             Technical Report, Institute of Mathematics, Russian Academy of Sciences, Ufa, 1991.
             
\bibitem{BSW} F. Beukers, J.A. Sanders, and J.P. Wang, One symmetry does not imply integrability.
             {\em J. Diff. Eq.} 146 (1998) 251--260.

\bibitem{Bilge} A.H. Bilge, A system with a recursion operator but one higher local symmetry.
              {\em Lie Groups Appl.} 1 (1994) 132--139. (Also available as
              arXiv:solv-int/9905012).

\bibitem{Ca} F. Calogero, Why Are Certain Nonlinear PDEs Both Widely Applicable and Integrable? In: 
            'What Is Integrability?', E. Zakharov (Ed.), 1-62.
             Springer-Verlag, Berlin Heidelberg 1991.
             
\bibitem{Calogero-2005}
F. Calogero, Isochronous Systems, Oxford University Press, Oxford, UK, 2008.

\bibitem{Calogero-Mariani-2005}
  F. Calogero and M. Mariani, A modified Schwarzian Korteweg-de Vries equation in 2+1 dimensions with lots of isochronous
   solutions,
  \textit{Phys. Atomic Nuclei} Vol. 68 (2005) 1646--1653; Russian version: Yadernaya Fiz. Vol. 68 (2005) 1710--1717.

\bibitem{Calogero-Euler-Euler}
  F. Calogero, M. Euler and N. Euler, New evolution PDEs with many isochronous solutions, \textit{J. Math. Anbal. Appl.}
  Vol. 353 (2009) 481--488.
  
\bibitem{CSS} F. Cooper, M. Shepard and P. Sodano, Solitary waves in a class of generalized
              Korteweg-de Vries equations. {\em Phys. Rev. E} 48 (1993), 4027--4032.

\bibitem{DS} D.K. Demskoy and V.V. Sokolov, On recursion operators for elliptic models. {\em Nonlinearity} 21 (2008),  1253--1264.

\bibitem{DK} B. Dey and A. Khare, Stability of compacton solutions. {\em Physical Review E} 58 (1998), R2741--R2744.
               
\bibitem{Fok} A. S. Fokas, A symmetry approach to exactly solvable evolution
              equations, {\em J. Math. Physics}, 21(6), (1980), 1318--1325.

\bibitem{Her1} R. Hern\'andez Heredero, Integrable Quasilinear Equations,
			{\em Teoret. Mat. Fiz.}, 133:2 (2002), 233--246.
			
\bibitem{Her2} R. Hern\'andez Heredero,
			Classification of Fully Nonlinear Integrable Evolution Equations of Third Order,
			{\em J. Nonlinear Math. Phys.}, 12  (2005), 567--585.
			
\bibitem{HHR} R. Hern\'andez Heredero and E.G. Reyes, Nonlocal symmetries, compacton equations and integrability.
              \textit{Int. J. Geometric Methods in Modern Physics} Vol. 10, No. 9 (2013) 1350046 (24 pages).

\bibitem{HSSv}
  R. Hern\'andez Heredero, V.V. Sokolov, S.I. Svinolupov,
  Why are there so many integrable equations of third order?, in {\it Proceedings of NEEDSâ94, Los Alamos}, 
  ed. E.V. Zhakarov, A.E. Bishop, D.D. Holm
  (World Scientific, 1994), pp. 42--53.

\bibitem{KC} A. Khare and F. Cooper, One-parameter family of soliton solutions with compact support in a class of generalized Korteweg-de Vries equations. {\em Physical Review E} 48 (1993), 4843--4844.

\bibitem{LD} A. Ludu and J. P. Draayer, Patterns on liquid surfaces: cnoidal
             waves, compactons and scaling, {\em Phys. D} 123 (1998), 82--91.

\bibitem{LOR} Yi.A Li, P.J. Olver, and P. Rosenau, Non-analytic solutions of nonlinear wave models.
               In `Nonlinear theory of generalized functions (Vienna, 1997)', 129--145.
               Chapman \& Hall/CRC, Boca Raton, FL, 1999.
               
\bibitem{Mariani-Calogero-2005}
  M. Mariani and  F. Calogero, Isochronous PDEs, \textit{Phys. Atomic Nuclei} Vol. 68 (2005) 899--908;
  Russian version: Yadernaya Fiz. Vol. 68 (2005) 935--944.
  
\bibitem{MSS}
  A.V. Mikhailov, A.B. Shabat, V.V. Sokolov,
  The symmetry approach to classification of integrable equations, in {\it What is Integrability?}, ed. E.V. Zhakarov
  (Springer, 1991), pp. 115--184.

\bibitem{MikSok} A.V. Mikhailov, V.V. Sokolov. Symmetries of Differential Equations and the Problem of Integrability. 
    Lect. Notes Phys. 767, (2009), 19--88.

\bibitem{O}  P.J. Olver, ``Applications of Lie Groups to Differential Equations'' (1993). Second
             Edition, Springer-Verlag, New York.
             
\bibitem{RZ} P. Rosenau and A. Zilburg, Compactons. {\em J. Phys. A: Math. Theor.} 51 (2018) 343001 (136pp).
             
\bibitem{SS} V.V. Sokolov, V.V. Shabat,
Classification of integrable evolution equations, \textit{Sov. Sci. Rev. C } \textbf{4} (1984) 221--280.

\bibitem{PNN} N. Petersson, N. Euler, M.  Euler, Recursion operators for a class of integrable third order equations.
              \textit{Studies in Applied Mathematics} \textbf{112} (2004), 201--225. 

\bibitem{Ro} P. Rosenau, What is $\dots$ a Compacton? {\em Notices of the AMS} 52 (2005), 738--739.

\bibitem{RH} P. Rosenau and J.M. Hyman, Compactons: Solitons with finite
             wavelength. {\em Phys. Rev. Lett.} 70 (1993), 564--567.
             
\bibitem{SviSok} S.I. Svinolupov, V.V. Sokolov. Weak nonlocalities in evolution equations. {\em Mathematical Notes}  48:6 (1990), 1234--1239.

\bibitem{SW} J.A. Sanders, J.-P. Wang, On the integrability of homogeneous scalar evolution
             equations. {\em J. Differential Equations} 147 (1998), 410--434.
             
\bibitem{SW1} J.A. Sanders and J.-P. Wang, On the integrability of
             non-polynomial scalar evolution equations. {\em J. Differential Equations} 166 (2000), 132--150.

\bibitem{Vo} J. Vodov\'a, A complete list of conservation laws for non-integrable compacton equations of $K(m,m)$ type. {\em Nonlinearity} 26 (2013), 757--762.

\bibitem{ZR} A. Zilburg and P. Rosenau, Loss of regularity in the $K(m,n)$ equations. {\em Nonlinearity} 31 (2018), 2651--2665.

\end{thebibliography}
\end{document}